\documentclass{nic-series}

\def\jnl@style#1{{\rmfamily#1}}%
\def\jref@jnl#1{{\jnl@style#1}}%

\newcommand\aj{\jref@jnl{AJ}}%
\newcommand\araa{\jref@jnl{ARA\&A}}%
\newcommand\apj{\jref@jnl{ApJ}}%
\newcommand\apjl{\jref@jnl{ApJ}}%
\newcommand\apjs{\jref@jnl{ApJS}}%
\newcommand\ao{\jref@jnl{Appl.~Opt.}}%
\newcommand\apss{\jref@jnl{Ap\&SS}}%
\newcommand\aap{\jref@jnl{A\&A}}%
\newcommand\aapr{\jref@jnl{A\&A~Rev.}}%
\newcommand\aaps{\jref@jnl{A\&AS}}%
\newcommand\azh{\jref@jnl{AZh}}%
\newcommand\baas{\jref@jnl{BAAS}}%
\newcommand\jcop{\jref@jnl{J.~Comp.~Phys.}}%
\newcommand\jkas{\jref@jnl{Journ.~Korean.~Astron.~Soc.}}%
\newcommand\jrasc{\jref@jnl{JRASC}}%
\newcommand\memras{\jref@jnl{MmRAS}}%
\newcommand\mnras{\jref@jnl{MNRAS}}%
\newcommand\na{\jref@jnl{New Astron.}}%
\newcommand\nar{\jref@jnl{New Astron. Rev.}}%
\newcommand\pra{\jref@jnl{Phys.~Rev.~A}}%
\newcommand\prb{\jref@jnl{Phys.~Rev.~B}}%
\newcommand\prc{\jref@jnl{Phys.~Rev.~C}}%
\newcommand\prd{\jref@jnl{Phys.~Rev.~D}}%
\newcommand\pre{\jref@jnl{Phys.~Rev.~E}}%
\newcommand\prl{\jref@jnl{Phys.~Rev.~Lett.}}%
\newcommand\pasa{\jref@jnl{PASA}}%
\newcommand\pasp{\jref@jnl{PASP}}%
\newcommand\pasj{\jref@jnl{PASJ}}%
\newcommand\qjras{\jref@jnl{QJRAS}}%
\newcommand\skytel{\jref@jnl{S\&T}}%
\newcommand\solphys{\jref@jnl{Sol.~Phys.}}%
\newcommand\sovast{\jref@jnl{Soviet~Ast.}}%
\newcommand\ssr{\jref@jnl{Space~Sci.~Rev.}}%
\newcommand\zap{\jref@jnl{ZAp}}%
\newcommand\nat{\jref@jnl{Nature}}%
\newcommand\iaucirc{\jref@jnl{IAU~Circ.}}%
\newcommand\aplett{\jref@jnl{Astrophys.~Lett.}}%
\newcommand\apspr{\jref@jnl{Astrophys.~Space~Phys.~Res.}}%
\newcommand\bain{\jref@jnl{Bull.~Astron.~Inst.~Netherlands}}%
\newcommand\fcp{\jref@jnl{Fund.~Cosmic~Phys.}}%
\newcommand\gca{\jref@jnl{Geochim.~Cosmochim.~Acta}}%
\newcommand\grl{\jref@jnl{Geophys.~Res.~Lett.}}%
\newcommand\jcp{\jref@jnl{J.~Chem.~Phys.}}%
\newcommand\jgr{\jref@jnl{J.~Geophys.~Res.}}%
\newcommand\jqsrt{\jref@jnl{J.~Quant.~Spec.~Radiat.~Transf.}}%
\newcommand\memsai{\jref@jnl{Mem.~Soc.~Astron.~Italiana}}%
\newcommand\nphysa{\jref@jnl{Nucl.~Phys.~A}}%
\newcommand\physrep{\jref@jnl{Phys.~Rep.}}%
\newcommand\physscr{\jref@jnl{Phys.~Scr}}%
\newcommand\planss{\jref@jnl{Planet.~Space~Sci.}}%
\newcommand\procspie{\jref@jnl{Proc.~SPIE}}%
\newcommand\znat{\jref@jnl{Z.~Naturforsch}}%

\begin{document} 

\title{Turbulence and its effect on protostellar disk formation}

\author{Daniel Seifried \inst{1} \and Robi Banerjee \inst{2} \and Ralf S. Klessen \inst{3}}

\institute{I. Physikalisches Institut,\\
         Universit\"at zu K\"oln, \\
         Z\"ulpicher Str. 77, 50937 K\"oln, Germany\\
         \email {seifried@ph1.uni-koeln.de}
         \and
         Hamburger Sternwarte, \\
         Universit\"at Hamburg, \\
         Gojenbergsweg 112, 21029 Hamburg, Germany\\
         \email{banerjee@hs.uni-hamburg.de}
         \and
         Universit\"at Heidelberg, Zentrum f\"ur Astronomie,\\
         Institut f\"ur Theoretische Astrophysik,\\
         Albert-Ueberle-Str. 2, 69120 Heidelberg, Germany \\
         \email{klessen@uni-heidelberg.de}
          }

\maketitle

\begin{abstracts}
We analyse simulations of turbulent, magnetised molecular cloud cores focussing on the formation of Class 0 stage protostellar discs and the physical conditions in their surroundings. We show that for a wide range of initial conditions Keplerian discs are formed in the Class 0 stage already. Furthermore, we show that the accretion of mass and angular momentum in the surroundings of protostellar discs occurs in a highly anisotropic manner, by means of a few narrow accretion channels. The magnetic field structure in the vicinity of the discs is highly disordered, revealing field reversals up to distances of 1000 AU. These findings demonstrate that as soon as even mild turbulent motions are  included, the classical disc formation scenario of a coherently rotating environment and a well-ordered magnetic field breaks down.
\end{abstracts}

\section{Introduction}
In our research we investigate the formation of protostars and their associated protostellar disks, the early precursors of planetary systems. During the last decade simulations of collapsing molecular cloud cores have revealed the so-called catastrophic magnetic braking problem: Magnetic fields are able to transport angular momentum by means of toroidal Alfv\`en waves. Modeling the collapse of rotating molecular cloud cores, simulations have shown that in the presence of magnetic fields with strengths comparable to observational results, the formation of rotationally supported (Keplerian) protostellar disks is largely suppressed \cite{Allen03,Hennebelle08}. This is due to the fact that angular momentum is removed very efficiently from the interior of the core by the magnetic field. In previous works we could confirm this effect for the collapse of massive (100 solar masses), molecular cloud cores \cite{Seifried11}. This key result of the suppression of Keplerian disk formation during the earliest stages of star formation is in contrast to recent observational results which state that protostellar disks should be present already in the Class 0 stage \cite{Tobin12,Sanchez13}.

\section{Numerical methods}

The simulations presented here are performed with the hydrodynamics code FLASH, which is written in Fortran 90 \cite{Fryxell00}. The code solves the 3-dimensional, discretized magnetohydrodynamical equations on a Cartesian grid. Making use of the adaptive-mesh-refinement (AMR) technique, only those regions which are of particular interest for us are resolved with the highest possible spatial resolution whereas other regions of minor interest are resolved more coarsely. This significantly reduces the number of calculations to be performed and hence the computational time required, thus allowing us to perform the simulations over long physical timescales. We also make use of the sink particle routine to model the formation of protostars \cite{Federrath10}.

\section{Initial conditions}

Observations of the birth places of stars show a wide  range of physical quantities, in particular in their initial mass. As we do not simulate a particular region observed by astronomers but rather aim to understand the systematic influence of the initial conditions, we have to perform a number of simulations in our work covering a wide range of masses and turbulence strengths. This allows us to draw conclusions about the effect of the initial conditions on the formation mechanism of stars. We modeled the collapse of molecular cloud cores with masses ranging from about 2 solar masses up to 1000 solar masses. The cores are threaded by a strong magnetic field along the z-axis and have an additional supersonic, turbulent velocity field as indicated by observations.

\section{Results}

The results of previous simulations described in section 1 show up in case that highly idealized initial conditions are used for the simulations. In particular the lack of turbulent motions - frequently observed in molecular cloud cores - could have a significant effect on the formation of protostellar disks and outflows. For this reason, in our research we here focus on the influence of turbulence on the formation of protostellar disks and outflows. This work has been performed on JUROPA and other supercomputing facilities. Each of the simulations required a computational time of a few 100 000 CPU-hours with a simultaneous use of up to 1000 CPUs per simulation. A few hundreds of files were produce for each simulation requiring a disk space of a few TB in total.

\subsection{Turbulence-induced disk formation}

We examine our simulations focusing on the question of how turbulence affects the formation of Keplerian disks \cite{Seifried12,Seifried13,Seifried15}. An example result is shown in Fig. 1 showing the protostellar disk in a representative run with a molecular cloud core of 100 solar masses.

\begin{figure}[t]
\begin{center}
\includegraphics[width=12cm]{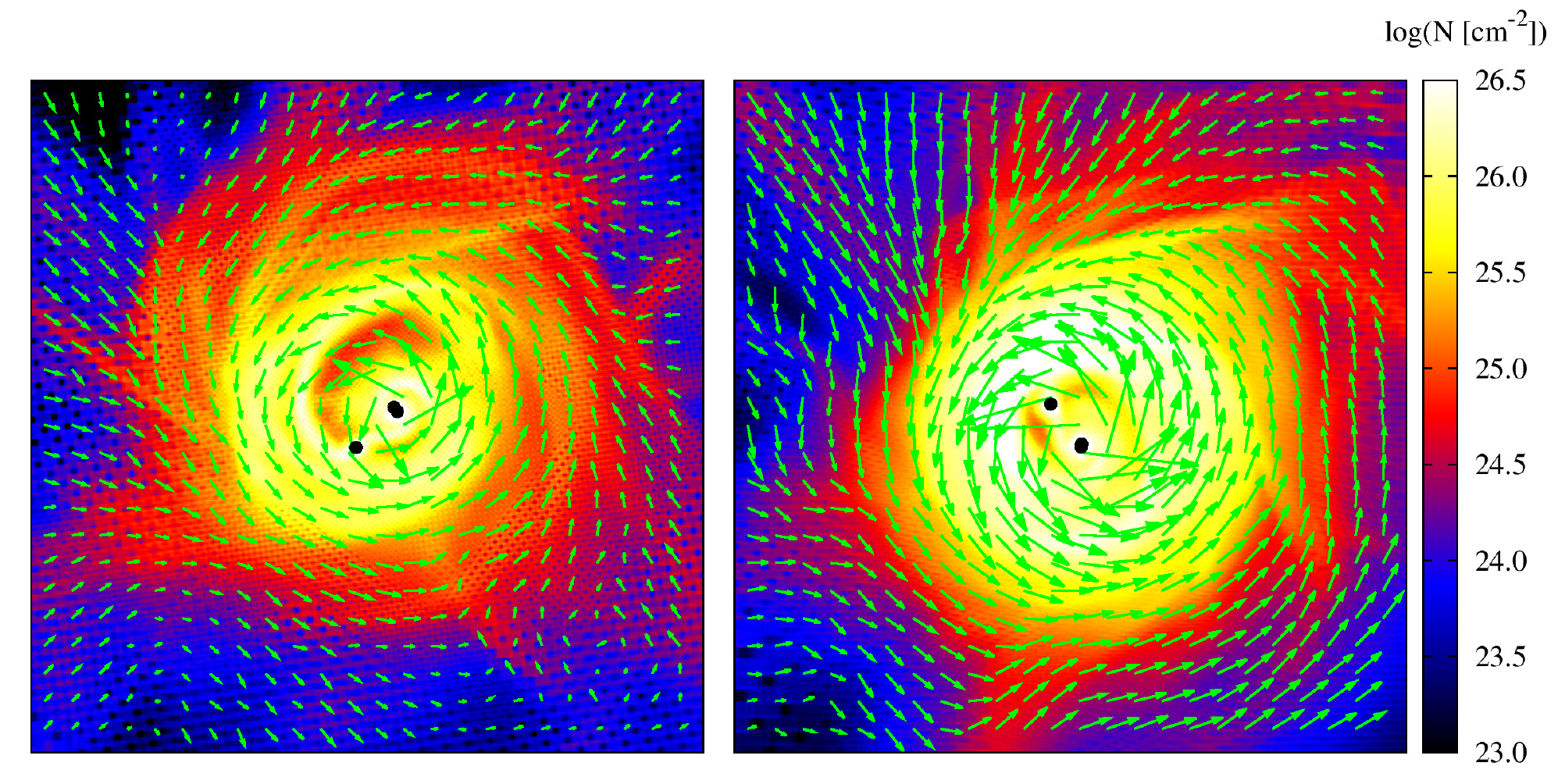}
\caption{Protostellar disk seen from top-on. Black dots represent protostars, green arrows the velocity field.}
\end{center}
\end{figure}
As can be seen, in the presence of turbulence rotationally supported disks are re-obtained again, which is in contrast to the previous simulations with comparable magnetic field strengths but no turbulence \cite{Seifried11}. This suggests that the efficiency of the magnetic braking, which is responsible for removing angular momentum from the midplane, is reduced significantly in the turbulent case. Analyzing the surroundings of the disks we can show that this indeed the case. The turbulent surroundings of the disk exhibit no coherent rotation structure (compare Fig. 1). Therefore, an efficient build-up of a strong toroidal magnetic field responsible for angular momentum extraction is hampered. Moreover, the turbulent motions lead to a strongly disordered magnetic field which further reduces the magnetic braking efficiency. Since simultaneously the angular momentum inwards transport remains high due to the presence of local shear flows in the vicinity of the disks, there is a net inwards angular momentum transport towards the center of the disk. The combination of these effects results in the observed build-up of Keplerian disks as expected from observations \cite{Tobin12,Sanchez13}. Varying the core masses (2.6 - 1000 solar masses) and the turbulence strengths  does not change our findings. This clearly demonstrate that the turbulence-induced disk formation mechanism works for a wide range of initial conditions. In particular, we could show that the formation of Keplerian disks does not require an uniform rotation of the core -- turbulent motions alone lead to the build-up of Keplerian disks. Moreover, we showed that even for subsonic turbulence, which is usually present in low-mass protostellar cores, the turbulence-induced formation mechanism still holds.

\begin{figure}[t]
\begin{center}
\includegraphics[width=12cm]{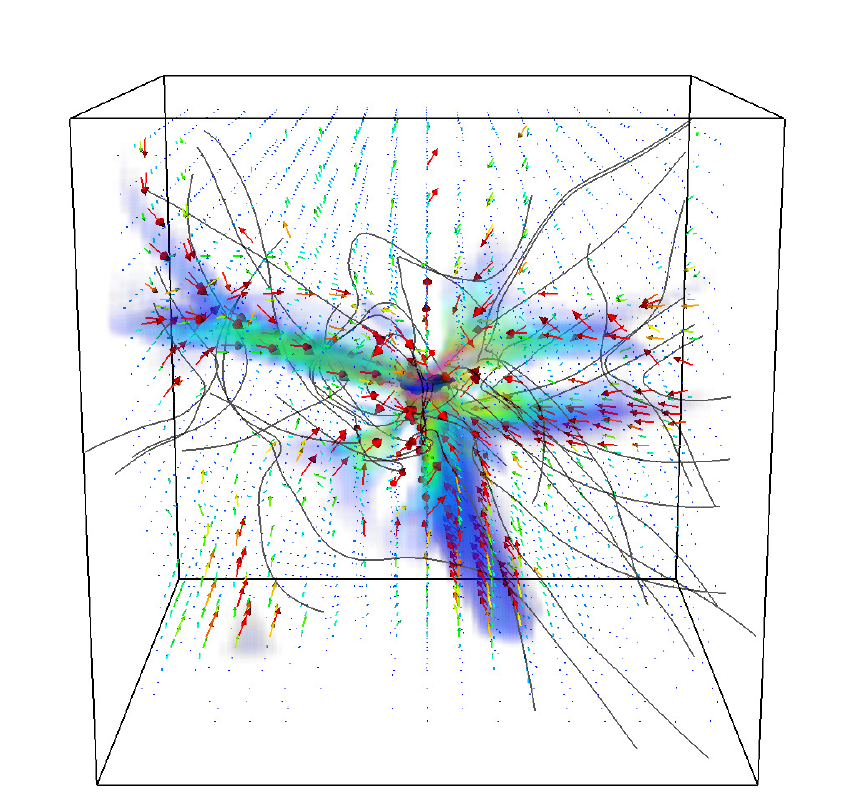}
\caption{3-dimensional structure of the magnetic field and gas motions around a Keplerian disk in one of our simulations.}
\end{center}
\end{figure}
In Fig. 2 we show the 3-dimensional structure of the magnetic field and the gas motions around a protostellar disk in one of our simulations. The magnetic field lines clearly reveal a highly complex structure being far off from well-ordered. Several field reversals up to distances of about 1000 AU from the disk center can be found. This indicates that the recently observed inclination of magnetic fields with respect to the disc axis \cite{Chapman13, Hull13,Hull14} could simply be the consequence of a spatially unresolved, highly disordered field structure.
The anisotropy of the accretion towards the disk is represented by the colored regions in the figure. Accretion is dominated by a few narrow accretion channels which carry a significant part of the inflowing mass and angular momentum but cover only about 10\% of the surface area. We emphasize that we find this anisotropic accretion mode in all of our simulations.

\section{Conclusions and outlook}

To summarize, our simulations present a richer picture of the process of disk formation, wherein turbulence, filamentary accretion streams, and magnetic field reversals guarantee that the otherwise overwhelming strength of magnetic braking by ordered fields is significantly degraded, allowing Keplerian discs to form. Non-turbulent collapse simulations, on the other hand, might significantly overestimate the efficiency of magnetic braking and thus underestimate the fraction of Class 0 stage Keplerian discs. We suggest that the anisotropic accretion and disordered magnetic field structure found in the environment of protostellar discs might set the stage for other mechanisms contributing to the formation of discs like (pseudo-) disc warping or non-ideal MHD effects \cite{Li14,Machida14}

For our future research we intend to study the self-consistent ejection of protostellar outflows from the Keplerian disks formed in our turbulence simulations. In order to reach this goal we will have to redo some of our simulations with increased spatial resolution, which will require further large amounts of computing power in the future. Furthermore, we plan to produce so-called synthetic observations. These synthetic observations will allow us to directly compare our simulation results with real observations. Such comparisons allow us to assess what can be inferred from observations - in particular how reliable parameters obtained from real observations are.

\section*{Acknowledgments}

DS acknowledges funding by the Bonn-Cologne Graduate School as well as the from the Deutsche Forschungsgemeinschaft (DFG) via the Sonderforschungsbereich SFB 956.
RB acknowledges funding by the DFG via the Emmy-Noether grant BA 3706/1-1, the ISM-SPP 1573 grants BA 3706/3-1 and BA 3706/3-2, as well as for the grant BA 3706/4-1.
RSK acknowledges support from the European Research Council under the European Communitys Seventh Framework Programme (FP7/2007-2013) via the ERC Advanced Grant STARLIGHT (project number 339177) and support from the DFG via the Sonderforschungsbereich SFB 881.
The FLASH code was developed partly by the DOE-supported Alliances Center for Astrophysical Thermonuclear Flashes (ASC) at the University of Chicago.

\bibliographystyle{nic}
\bibliography{literature}

\begin{thebibliography}{10}

\bibitem{Allen03}
A.~{Allen}, {Z.-Y.} {Li}, and F.~H. {Shu},
 {\em {Collapse of Magnetized Singular Isothermal Toroids. II. Rotation and
  Magnetic Braking}},
 \apj, {\bf 599}, 363--379, Dec. 2003.

\bibitem{Hennebelle08}
P.~{Hennebelle} and S.~{Fromang},
 {\em {Magnetic processes in a collapsing dense core. I. Accretion and
  ejection}},
 \aap, {\bf 477}, 9--24, Jan. 2008.

\bibitem{Seifried11}
D.~{Seifried}, R.~{Banerjee}, R.~S. {Klessen}, D.~{Duffin}, and R.~E.
  {Pudritz},
 {\em {Magnetic fields during the early stages of massive star formation - I.
  Accretion and disc evolution}},
 \mnras, {\bf 417}, 1054--1073, Oct. 2011.

\bibitem{Tobin12}
J.~J. {Tobin}, L.~{Hartmann}, H.-F. {Chiang}, D.~J. {Wilner}, L.~W. {Looney},
  L.~{Loinard}, N.~{Calvet}, and P.~{D'Alessio},
 {\em {A \~{}0.2-solar-mass protostar with a Keplerian disk in the very young
  L1527 IRS system}},
 \nat, {\bf 492}, 83--85, Dec. 2012.

\bibitem{Sanchez13}
{\'A}.~{S{\'a}nchez-Monge}, R.~{Cesaroni}, M.~T. {Beltr{\'a}n}, M.~S.~N.
  {Kumar}, T.~{Stanke}, H.~{Zinnecker}, S.~{Etoka}, D.~{Galli}, C.~A. {Hummel},
  L.~{Moscadelli}, T.~{Preibisch}, T.~{Ratzka}, F.~F.~S. {van der Tak},
  S.~{Vig}, C.~M. {Walmsley}, and K.-S. {Wang},
 {\em {A candidate circumbinary Keplerian disk in G35.20-0.74 N: A study with
  ALMA}},
 \aap, {\bf 552}, L10, Apr. 2013.

\bibitem{Fryxell00}
B.~{Fryxell}, K.~{Olson}, P.~{Ricker}, F.~X. {Timmes}, M.~{Zingale}, D.~Q.
  {Lamb}, P.~{MacNeice}, R.~{Rosner}, J.~W. {Truran}, and H.~{Tufo},
 {\em {FLASH: An Adaptive Mesh Hydrodynamics Code for Modeling Astrophysical
  Thermonuclear Flashes}},
 \apjs, {\bf 131}, 273--334, Nov. 2000.

\bibitem{Federrath10}
C.~{Federrath}, R.~{Banerjee}, P.~C. {Clark}, and R.~S. {Klessen},
 {\em {Modeling Collapse and Accretion in Turbulent Gas Clouds: Implementation
  and Comparison of Sink Particles in AMR and SPH}},
 \apj, {\bf 713}, 269--290, Apr. 2010.

\bibitem{Seifried12}
D.~{Seifried}, R.~{Banerjee}, R.~E. {Pudritz}, and R.~S. {Klessen},
 {\em {Disc formation in turbulent massive cores: circumventing the magnetic
  braking catastrophe}},
 \mnras, {\bf 423}, L40--L44, June 2012.

\bibitem{Seifried13}
D.~{Seifried}, R.~{Banerjee}, R.~E. {Pudritz}, and R.~S. {Klessen},
 {\em {Turbulence-induced disc formation in strongly magnetized cloud cores}},
 \mnras, {\bf 432}, 3320--3331, July 2013.

\bibitem{Seifried15}
D.~{Seifried}, R.~{Banerjee}, R.~E. {Pudritz}, and R.~S. {Klessen},
 {\em {Accretion and magnetic field morphology around Class 0 stage
  protostellar discs}},
 \mnras, {\bf 446}, 2776--2788, Jan. 2015.

\bibitem{Chapman13}
N.~L. {Chapman}, J.~A. {Davidson}, P.~F. {Goldsmith}, M.~{Houde}, W.~{Kwon},
  Z.-Y. {Li}, L.~W. {Looney}, B.~{Matthews}, T.~G. {Matthews}, G.~{Novak},
  R.~{Peng}, J.~E. {Vaillancourt}, and N.~H. {Volgenau},
 {\em {Alignment between Flattened Protostellar Infall Envelopes and Ambient
  Magnetic Fields}},
 \apj, {\bf 770}, 151, June 2013.

\bibitem{Hull13}
C.~L.~H. {Hull}, R.~L. {Plambeck}, A.~D. {Bolatto}, G.~C. {Bower}, J.~M.
  {Carpenter}, R.~M. {Crutcher}, J.~D. {Fiege}, E.~{Franzmann}, N.~S.
  {Hakobian}, C.~{Heiles}, M.~{Houde}, A.~M. {Hughes}, K.~{Jameson}, W.~{Kwon},
  J.~W. {Lamb}, L.~W. {Looney}, B.~C. {Matthews}, L.~{Mundy}, T.~{Pillai},
  M.~W. {Pound}, I.~W. {Stephens}, J.~J. {Tobin}, J.~E. {Vaillancourt}, N.~H.
  {Volgenau}, and M.~C.~H. {Wright},
 {\em {Misalignment of Magnetic Fields and Outflows in Protostellar Cores}},
 \apj, {\bf 768}, 159, May 2013.

\bibitem{Hull14}
C.~L.~H. {Hull}, R.~L. {Plambeck}, W.~{Kwon}, G.~C. {Bower}, J.~M. {Carpenter},
  R.~M. {Crutcher}, J.~D. {Fiege}, E.~{Franzmann}, N.~S. {Hakobian},
  C.~{Heiles}, M.~{Houde}, A.~M. {Hughes}, J.~W. {Lamb}, L.~W. {Looney}, D.~P.
  {Marrone}, B.~C. {Matthews}, T.~{Pillai}, M.~W. {Pound}, N.~{Rahman},
  G.~{Sandell}, I.~W. {Stephens}, J.~J. {Tobin}, J.~E. {Vaillancourt}, N.~H.
  {Volgenau}, and M.~C.~H. {Wright},
 {\em {TADPOL: A 1.3mm Survey of Dust Polarization in Star-forming Cores and
  Regions}},
 \apjs, {\bf 213}, 13, July 2014.

\bibitem{Li14}
Z.-Y. {Li}, R.~{Krasnopolsky}, H.~{Shang}, and B.~{Zhao},
 {\em {On the Role of Pseudodisk Warping and Reconnection in Protostellar Disk
  Formation in Turbulent Magnetized Cores}},
 ArXiv e-prints, Aug. 2014.

\bibitem{Machida14}
M.~N. {Machida}, S.-i. {Inutsuka}, and T.~{Matsumoto},
 {\em {Conditions for circumstellar disc formation: effects of initial cloud
  configuration and sink treatment}},
 \mnras, {\bf 438}, 2278--2306, Mar. 2014.

\end{thebibliography}

\end{document}